\documentclass[aps,pre,twocolumn,reprint,10pt,superscriptaddress,eprint,longbibliography]{revtex4-2}
\usepackage{libertinus}
\usepackage{graphicx}
\usepackage[hidelinks]{hyperref}
\usepackage{orcidlink}
\usepackage[capitalise]{cleveref}
\usepackage{microtype}
\usepackage{xcolor}
\hypersetup{
    colorlinks,
    linkcolor={red!50!black},
    citecolor={blue!50!black},
    urlcolor={blue!80!black}
}

\begin{document}
\title
{
Learning minimal representations of stochastic processes\\
with variational autoencoders
}

\author{Gabriel Fern\'andez-Fern\'andez\,\orcidlink{0000-0001-6719-8275}}%
\affiliation{ICFO – Institut de Ci\`encies Fot\`oniques, The Barcelona Institute of Science and Technology,
Av. Carl Friedrich Gauss 3, 08860 Castelldefels (Barcelona), Spain}

\author{Carlo Manzo\,\orcidlink{0000-0002-8625-0996}}
\affiliation{%
Facultat de Ci\`encies, Tecnologia i Enginyeries, Universitat de Vic – Universitat
Central de Catalunya (UVic-UCC), C. de la Laura, 13, 08500 Vic, Spain
}
\affiliation{
Institut de Recerca i Innovaci\'o en Ci\`encies de la Vida i de la Salut a la Catalunya Central (IRIS-CC), 08500 Vic, Barcelona, Spain
}
\author{Maciej Lewenstein\,\orcidlink{0000-0002-0210-7800}}
\affiliation{ICFO – Institut de Ci\`encies Fot\`oniques, The Barcelona Institute of Science and Technology,
Av. Carl Friedrich Gauss 3, 08860 Castelldefels (Barcelona), Spain}
\affiliation{ICREA, Pg. Llu\'is Companys 23, 08010 Barcelona, Spain}

\author{\\Alexandre Dauphin\,\orcidlink{0000-0003-4996-2561}}
\affiliation{ICFO – Institut de Ci\`encies Fot\`oniques, The Barcelona Institute of Science and Technology,
Av. Carl Friedrich Gauss 3, 08860 Castelldefels (Barcelona), Spain}%

\author{Gorka Mu\~noz-Gil\,\orcidlink{0000-0001-9223-0660}}%
\affiliation{ 
Institute for Theoretical Physics, University of Innsbruck, Technikerstr. 21a, A-6020 Innsbruck, Austria
}%

\begin{abstract}
Stochastic processes have found numerous applications in science, as they are broadly used to model a variety of natural phenomena.  Due to their intrinsic randomness and uncertainty, they are, however, difficult to characterize.  Here, we introduce an  unsupervised machine learning approach  to determine the minimal set of parameters required to effectively describe the dynamics of a stochastic process. Our method builds upon
an extended $\beta$-variational autoencoder architecture. By means of simulated datasets corresponding to paradigmatic diffusion models, we showcase its effectiveness in extracting the minimal relevant parameters that accurately describe these dynamics. Furthermore, the method enables the generation of new trajectories that faithfully replicate the expected stochastic behavior. 
Overall, our approach enables the autonomous discovery of unknown parameters describing stochastic processes, hence enhancing our comprehension of complex phenomena across various fields.
\end{abstract}

\keywords{Unsupervised, Machine Learning, Variational Autoencoder, Feature Learning, Representation Learning, Autoregressive Model, Data Analysis, Knowledge Discovery, Interpretability, Anomalous Diffusion, Single Trajectory Characterization, Time Series, Stochastic Model, Brownian Motion, Fractional Brownian Motion, Scaled Brownian Motion}
\maketitle

\paragraph*{Introduction ---} 
The recent advances in machine learning (ML) have not only impacted everyday life but also the development of science. In physics, the predictive power of ML has been used to get insights from theoretical and experimental physical systems with unprecedented accuracy~\cite{carleo2019machine, dawid2022modern}. Indeed, ML can easily extract knowledge from a plethora of data types with no prior information about its source. 

It has broadly been argued that, if a machine can make predictions over a given physical process, the properties of the latter must be encoded in the internal representation of the machine~\cite{molnar2018interpretable}. Therefore, beyond its predictive nature, ML can also be helpful for scientific discovery. Several examples in
biology~\cite{soelistyo2022learning},
quantum matter~\cite{miles2021correlator, kaming2021unsupervised}, quantum information~\cite{pozas2023proofs}, lattice field theory~\cite{blucher2020novel}, mathematics~\cite{davies2021advancing}, or experiment design~\cite{melnikov2018active,krenn2020computer} show that deep neural networks (NN), despite being often considered as black boxes,  can guide scientists to understand complex phenomena or to design involved experiments. 

Various techniques exploit the information encoded in the trained model, e.g., by defining a notion of similarity between the different training examples and the test examples~\cite{koh2017understanding, dawid2021hessian}. Alternatively, the study of the internal representation of the NN allows mapping the statistics of the training and test examples onto a vectorial space. An example of such embedding is the encoding produced at the bottleneck layer of an autoencoder (AE)~\cite{hinton2006reducing}, NN architectures trained to compress and decompress data to and from a given vectorial space. The abstract representation obtained at this level is very useful for unsupervised applications and machine interpretability. For instance, AE representations, and in particular those of variational autoencoders (VAE)~\cite{kingma2013auto, rezende2014stochastic}, are currently crucial for the efficient training of some of the most powerful ML models, from image generation with diffusion models~\cite{rombach2022high} to reinforcement learning~\cite{higgins2017darla}. 
Beyond such downstream applications, focused on enhancing the power of ML methods, VAEs can be used to discover hidden factors of variation in an unsupervised way~\cite{higgins2016beta, burgess2018understanding, chen2018isolating}. These models create disentangled representations and isolate the generating factors of input datasets. Achieving this objective is not trivial and great effort is currently invested into improving such representations~\cite{locatello2019challenging, eddahmani2023unsupervised}.
This application is particularly interesting in physical systems, as the generating factors translate to the relevant physical parameters of the system. Their utility has been extensively proven in a variety of scenarios, in particular in the analysis  of dynamical systems~\cite{iten2020discovering, lu2020extracting}. These seminal works mainly focused on deterministic systems, raising questions about their applicability to a wider range of real-world scenarios, especially those involving stochastic processes.
Similar approaches have also been proposed for stochastic processes (see e.g.,~\cite{choi2022learning, im2023data}), but they rely on the preprocessing/averaging of the data, such that the input to the ML model is effectively a deterministic signal. Therefore, the efficiency of the representation achieved depends not only on the accuracy of the ML model but also on the statistical relevance of its inputs.
Thus, it would be beneficial to develop models that can treat raw stochastic data. This  kind of approach has been explored by several ML methods with great success. However, these methods rely on prior information about the system such as, e.g., the underlying physical model~\cite{frishman2020learning, bruckner2020inferring} or a basis of preselected functions~\cite{huang2022sparse, both2021deepmod}. Similarly, symbolic regression approaches can find the governing equations of stochastic processes~\cite{la2022distilling} but necessitate a meaningful set of initial terms to build a proposed expression. This further impedes their application to systems involving parameters that cannot be expressed in closed form.

In this work, we aim at determining whether a machine can extract, in an unsupervised way, the minimal parametric representation of a stochastic process from trajectories without any prior knowledge of the system. Extending previous works on unsupervised learning approaches to diffusion~\cite{kabbech2022identification, munoz2021unsupervised}, we train a $\beta$-variational autoencoder ($\beta$-VAE)~\cite{higgins2016beta} to generate trajectories with the same properties as the ones used as inputs. The architecture presents an information bottleneck constructed to represent conditionally independent factors of variation. Using an adaption of the original $\beta$-VAE~\cite{chorowski2019unsupervised}, we successfully train the architecture with various sets of data corresponding to diffusion processes with different characteristics. Our results show that only the minimal necessary properties describing the motion of the particles arise in the bottleneck and can be directly related to the known theories describing these models. Moreover, the training provides a generative model that can produce new trajectories with the same properties as the training dataset, thus allowing for an in-depth study of their statistical properties. Besides its fundamental value, this work offers a valuable tool for the study of molecular diffusion from individual trajectories, such as those obtained with single-molecule imaging techniques~\cite{manzo2015review, barkai2012strange}, for which extensive ML methods have been developed~\cite{munoz2021objective}. In contrast to the latter, rather than predicting known parameters with increasing accuracy, we aim here at solving a more fundamental question: learning the most efficient description of a given stochastic process.

\begin{figure}
    \includegraphics[width=\linewidth]{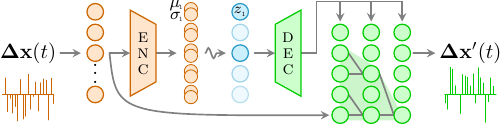}
    \caption{\label{fig:arch}
    \textbf{Interpretable autoregressive $\beta$-VAE.}
    Given the displacements $\mathbf{\Delta x}(t)$ of a diffusion trajectory, the encoder (orange) compresses them into an interpretable latent space (blue), in which few neurons (dark blue) represent physical features of the input data while others are noised out (light blue). An autoregressive decoder (green) generates from this latent representation the displacements $\Delta \mathbf{x'}(t)$ of a new trajectory recursively, considering a certain receptive field RF (light green cone).
    }
\end{figure}

\paragraph*{Interpretable generative model ---} 
We aim to construct a machine learning (ML) architecture capable of (i) extracting interpretable physical variables from stochastic time series, and (ii) modeling the probability distribution function of the input data.
To this end, we consider a $\beta$-variational autoencoder ($\beta$-VAE) architecture~\cite{higgins2016beta}, schematically depicted in \cref{fig:arch} (see \cref{sec:arch} and Ref.~\cite{fernandez2023spivae} for further details). In this architecture, an encoder (depicted in orange in \cref{fig:arch}) compresses displacements from an input trajectory $\mathbf{x}$ into a latent space $\mathbf{z}$ (shown in blue), for which each neuron is parameterized via a normal distribution $\mathcal{N}(\mu_{z_i}, \sigma_{z_i})$. Throughout this work, we consider $|\mathbf{z}|=6$ latent neurons. A sample is then drawn from the latent space and fed into the decoder (depicted in green), which generates a distribution function from which the displacements of new trajectories $\mathbf{x'}$ can be sampled.
The training of this architecture is based on a loss function that consists of two terms: a reconstruction loss that compares the model's inputs and outputs, and a second loss term that measures the dissimilarity between the distribution of the latent variables and their prior. For the prior, a standardized normal distribution $\mathcal{N}(0,1)$ is typically considered.
A parameter $\beta$ is used to control the relative weight of the two loss components and, through an \emph{ad hoc} annealing schedule, can be tuned in such a way that only the minimum number of latent neurons remains informative (i.e., $\sigma_{z_i} \ll 1)$. In a physical context, only the pertinent properties governing the process will manifest in the latent space and serve to reproduce the input~\cite{iten2020discovering}.

Traditional approaches for representation learning typically focus on deterministic decoders and use a reconstruction error to directly compare the input and output of the autoencoder~\cite{wetzel2017unsupervised, iten2020discovering}.
However, when dealing with stochastic signals, any form of compression inevitably leads to significant information loss in the reconstructed trajectory.
Therefore, we adopt a distinct approach based on a probabilistic decoder to model the distribution of displacements $p(\mathbf{\Delta x})$ through $p_\theta(\mathbf{\Delta x} | \mathbf{z})$, where $\theta$ represents the trainable parameters from which the individual displacements $\Delta x_i$ of the trajectories are sampled.
We thus use the maximum likelihood estimation to compare the resulting $p_\theta$ with the samples of the training dataset, assumed to be representative of the distribution $p$.
Notably, the stochastic signal that corresponds to the input may exhibit various types of correlations, which play a crucial role in modeling important physical processes.
To ensure that these properties are preserved at the autoencoder output, we follow the architecture proposed in Ref.~\cite{chorowski2019unsupervised} and construct the decoder using an autoregressive (AR) convolutional network known as WaveNet~\cite{oord2016wavenet}.
WaveNet models the output distribution according to the following recursive conditional probability
\begin{equation}
\label{eq:cond_prob}
p_\theta(\mathbf{\Delta x}|\mathbf{z})=\prod_{t = 1}^{T} p_\theta(\Delta x_t|\Delta x_{t-1},..., \Delta x_{t-\mathrm{RF}}; \mathbf{z}),
\end{equation}
where $T$ is the length of the trajectory and RF is the receptive field, i.e., the number of past displacements used to predict the forthcoming one (light green cone in \cref{fig:arch}).

\begin{figure}
    \centering
    \includegraphics[width=\linewidth]{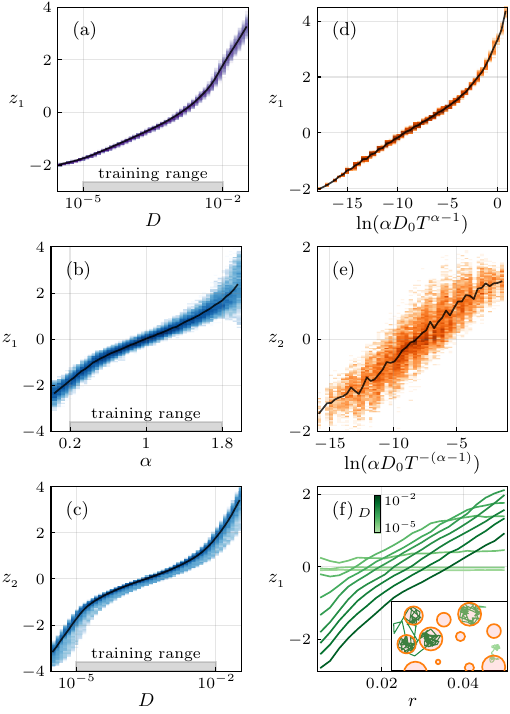}%
    \caption{
    \textbf{Interpretation of the latent space.} 
    Distribution of latent neuron activations \(z_i\) for four datasets: (a)~BM, (b, c)~FBM, (d, e)~SBM, and (f)~BM with confinement. Only surviving neurons are shown (i.e., $\sigma_{z_i} \ll 1$). For all datasets, the number of surviving neurons agrees with their respective number of degrees of freedom.  
    }
    \label{fig:latent}
\end{figure}

\paragraph*{Extracting physical variables from stochastic data ---}
To test the ability of the architecture to extract relevant physical variables from stochastic data, we train it on four datasets, constructed with three paradigmatic models of diffusion. First, we consider Brownian motion (BM)~\cite{einstein1905uber, von1906kinetischen}, used to describe the stochastic motion of a particle suspended in a fluid. 
The diffusion of a Brownian particle is characterized by a single parameter, the diffusion coefficient $D$, hence serving as an initial benchmark for our study. 
More precisely, BM can be expressed as a Langevin equation of the form $\mathbf{\dot{x}}(t) = \xi(t)$, where $\xi(t)$ is a Gaussian noise with autocorrelation function
\begin{equation}
\langle \xi(t)\xi(t') \rangle = 2D \delta(t-t').
\end{equation}%
To train the autoencoder, we generate a dataset of trajectories with $D\in [10^{-5}, 10^{-2}]$ using the \texttt{andi\_datasets} library~\cite{munoz2023andichallenge}.
As training proceeds, the model improves its generative capabilities while minimizing the KL divergence of the latent neurons with respect to their prior $\mathcal{N}(0,1)$. After training, a single neuron of the six available \textit{survives}, i.e., differs drastically from its purely noisy prior.
\Cref{fig:latent}(a) shows a direct relation between such neuron and $D$, highlighting that the autoencoder has learned that the only information needed by the decoder to generate a new trajectory is its diffusion coefficient. Furthermore, the correlation between $z_1$ and $D$ extends beyond the training set range (gray shaded area), indicating the model's ability to generalize the representation of diffusive parameters beyond the specified training range.

Next, we consider two extensions of BM, namely fractional Brownian motion (FBM), and scaled Brownian motion (SBM). These are paradigmatic models of anomalous diffusion, i.e., diffusion that deviates from the typical Brownian behavior. 
These models have found extensive application in describing motion in different biological scenarios at various scales~\cite{sabri2020elucidating,lim2002self, munoz2022stochastic, wang2022anomalous} and thus constitute a valuable benchmark to demonstrate the method's utility in experimental settings.
Both models are characterized by only two parameters: the diffusion coefficient and the anomalous diffusion exponent $\alpha$. However, the source of anomalous diffusion is different in each model.  

FBM can be derived from the Langevin equation and expressed as  $\mathbf{\dot{x}}(t) = \xi_{fGn}(t)$, where $\xi_{fGn}(t)$ represents fractional Gaussian noise with the autocorrelation function
\begin{equation}
\label{eq:fbm}
\langle \xi_{fGn}(t)\xi_{fGn}(t') \rangle = \alpha(\alpha-1)D \left|t-t'\right|^{\alpha-2}, 
\end{equation}
where $D$ is here referred to as a generalized diffusion coefficient with dimensions $[l]^2 [t]^{-{\alpha}}$. Importantly, \cref{eq:fbm} implies that FBM displacements are correlated. This feature provides an interesting benchmark for the autoregressive properties of the decoder, as we will discuss in the following section.  We train an autoencoder with a dataset consisting of FBM trajectories with $\alpha \in [0.2, 1.8]$ and $D\in [10^{-5}, 10^{-2}]$. In \cref{fig:latent}(b,c), we show the only two surviving neurons: one ($z_1$) shows a nearly linear relation with the anomalous diffusion exponent $\alpha$, whereas the other ($z_2$) has a monotonic dependence on the $\log(D)$. These results prove the model's ability to only retain minimal information to correctly reproduce FBM trajectories through the probabilistic decoder.

SBM extends Brownian diffusion by considering an aging diffusion coefficient $D(t)$, which is usually considered to scale as $D_\alpha(t)=\alpha D_{0} t^{\alpha-1}$, where $\alpha$ is the anomalous diffusion exponent and $D_0$ is a constant with dimensions $[l]^2 [t]^{-{\alpha}}$. After training, we again observed that only two neurons survived. However, in contrast to earlier cases, these two neurons exhibit a more intricate relationship with $\alpha$ and $D_0$, as depicted in \cref{fig:latent}(d,e). 
It must be pointed out that the only constraint imposed by the $\beta$-VAE loss in order to obtain these results is that the representation in the latent space is minimal, while still achieving good reconstruction loss. Hence, nothing prevents the network to learn a minimal representation based on combinations of the independent factors (meaningful physical variables in this case)~\cite{scholkopf2021toward, nautrup2022operationally}. Nonetheless, the number of surviving neurons should never exceed the number of independent factors (or degrees of freedom), a situation that would not correspond to a minimal representation.

In many scenarios, the factors of variation may not be inherently linked to a closed-form equation, unlike the examples previously mentioned. In such cases, phenomenological models are frequently employed to accurately capture the dynamics of the system by defining a few relevant parameters. The methodology we propose is particularly useful for deriving these parameters in situations where no prior information about the physical process exists. As an example,  we examine the behavior of a Brownian particle subjected to random confinement, a phenomenon prevalent in numerous biological contexts~\cite{sugiyama2023confinement, torreno2014enhanced}. This scenario involves a particle freely diffusing in a medium containing circular compartments of random sizes, where the particle reflects off the boundaries with a certain probability (\cref{fig:latent}(f) inset and \cref{sec:datasets}). Each particle's behavior can be characterized by two factors: its diffusion coefficient $D$ and, if applicable, the confinement radius $r$ of the compartment it enters. Unlike $D$ and $\alpha$, identifying $r$ poses a greater challenge. Given the random radius distribution and the boundaries' partial transmittance, it is necessary to isolate the confined segments within the overall trajectory and then compute their confinement radius~\cite{monnier2015inferring, argun2021classification}. Remarkably, the autoencoder is able to overcome such challenge and not only identifies this factor (\cref{fig:latent}(f)), but achieves it with remarkable precision. This suggest that the encoder has learned to segment the input trajectory autonomously and derive $r$ without any supervised guidance. For sufficiently large values of $D$, the surviving neuron correlates with the confinement radius. Expectedly, for smaller $D$ values, the particle's motion within the compartment is too limited, making the accurate determination of $r$ unfeasible.

\paragraph*{Generating trajectories from meaningful representations ---}
An essential feature of the presented architecture is its ability to generate trajectories with the same physical properties as the training samples. Moreover, the representational power of the latent space allows one to set the properties of the output trajectories by tuning the value of the latent neurons. Inference is then done directly from the latent space, without any need of the encoder.

As expressed in \cref{eq:cond_prob}, the decoder predicts the probability of each displacement $\Delta x_t$ by means of a conditional probability related to previous displacements and, most importantly, the latent vector $\mathbf{z}$. In practice, by means of the reparameterization trick~\cite{kingma2013auto}, the decoder outputs the mean $\mu_t$ and variance $\sigma_t^2$ of a normal distribution $\mathcal{N}(\mu_t, \sigma_t^2)$, and we then use the latter to sample each displacement $\Delta x_t$. In the case of BM trajectories, the autoencoder correctly learns to set $\mu_t = 0$ and $\sigma_t^2 = 2D \ \forall \ t$, as the displacements of such trajectories are independent and stationary. Hence, the decoder only needs to properly learn the exact transformation from $z_1$ in \cref{fig:latent}(a) to $\sigma_t^2$.

\begin{figure}
    \centering
    \includegraphics[width=\linewidth]{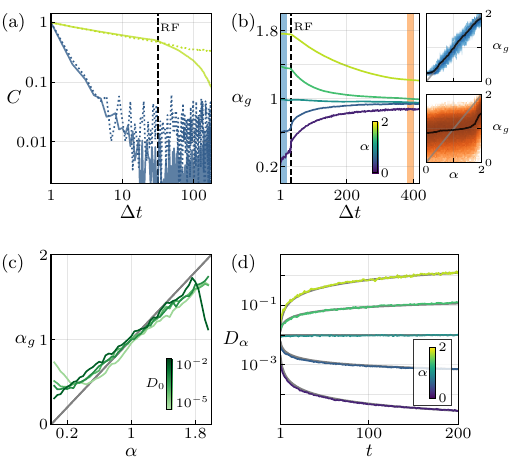}%
    \caption{\label{fig:generation}
    \textbf{Statistical properties of generated anomalous diffusion.}
    Top (bottom) row corresponds to the FBM (SBM) dataset.
    \textbf{(a)} Displacements correlations $C=\left|\langle \Delta x_t \Delta x_{t+\Delta t} \rangle\right|/\Delta x_0^2$ for the input (dotted) and generated data (solid) with $\alpha=0.6, 1.8$ (blue and green, respectively).
    \textbf{(b)} Anomalous exponent $\alpha_g$ of the generated FBM data fitted from the time-averaged mean squared displacement at different $\Delta t$ for different input $\alpha$.
    Insets show the two-dimensional histograms of the input vs. generated anomalous exponent at the highlighted $\Delta t$, before (blue) and after (orange) the receptive field RF.
    \textbf{(c)} Anomalous exponent $\alpha_g$ of the generated SBM data vs. the input exponent $\alpha$ for various $D_0$.
    \textbf{(d)} Evolution of the diffusion coefficient for generated SBM trajectories at various $\alpha$. Dotted lines show the expected scalings.
    }
\end{figure}

Next, we analyze the more complex cases of FBM and SBM. In this sense, a fundamental feature of FBM trajectories is the correlation of displacements, which has a characteristic power-law behavior directly connected to \cref{eq:fbm}. As commented, the architecture includes an autoregressive decoder to preserve this feature in generated trajectories. In fact, in \cref{fig:generation}(a), we show that when generating trajectories for a given $\alpha$, the power-law correlation is preserved in a range defined by the architecture's receptive field (RF) and then lost, as expected from \cref{eq:cond_prob}. Since power-law correlations produce anomalous diffusion in FBM, their loss affects the anomalous diffusion exponent of the generated trajectories, as shown in \cref{fig:generation}(b) (see \cref{sec:loss} for details). While the exponent is correct for $\Delta t < RF$, it rapidly converges to one at longer times. In our experiments, increasing the RF hindered training substantially. A possible solution is to consider a transformer-based decoder~\cite{vaswani2017attention}, where extensive efforts to enlarge context length are currently being pursued~\cite{bulatov2022recurrent}. 

With respect to the SBM dataset, the $\beta$-VAE must encode into the latent space the time-dependent diffusion coefficient $D_\alpha(t)$ in order to generate trajectories with anomalous diffusion exponent $\alpha$. We have shown that the latent space obtained for the model trained on SBM trajectories has a complex relationship with the input parameters $\alpha$ and $D_0$. To simplify the analysis, instead of generating trajectories directly from the latent space as we did with FBM, we feed trajectories with a given ground-truth $\alpha$ and $D_0$ to the encoder, extract their latent representation $\mathbf{z}$, and use it to generate new trajectories. As shown in \cref{fig:generation}(c), the generator is able to correctly reproduce trajectories with the correct exponent for various $D_0$ and a wide range of $\alpha$. In \cref{fig:generation}(d), we show $D_\alpha(t)$ calculated as the variance of the displacements for different $t$. The $\beta$-VAE perfectly reproduces the expected behavior over all generated times, confirming the generative capabilities of the architecture.

\paragraph*{Conclusions ---}
In this work, we have explored the application of machine learning (ML) techniques to provide interpretable representations of stochastic processes from time series. We have shown that a method based on a $\beta$-variational autoencoder with an autoregressive decoder can retrieve the minimal parametric representation of trajectories corresponding to different processes describing diffusion.

The architecture has been specially developed to account for common features present in stochastic data. First, the output of the network is probabilistic. Due to the stochastic nature of diffusion trajectories, trajectory reconstruction after compression is effectively unfeasible. Hence, instead of reconstructing, as done typically in AE, we aim at generating new trajectories via a parameterized distribution optimized to match the input data distribution. Second, the decoder is autoregressive, a feature introduced in order to model distributions with correlations, as for the case of FBM trajectories.  

In contrast to the predominantly employed supervised methods, our study showcases the potential of unsupervised machine learning techniques to uncover the intrinsic structure of stochastic processes and determine the minimal parameterization required for accurate characterization.
As such, it offers a promising avenue for uncovering previously unknown physical degrees of freedom inherent in stochastic physical processes. A significant advantage of this approach is its ability to operate without prior information about the data or the underlying physical process. This makes it particularly well-suited for experimental settings. In this regard, a promising avenue is the one related to \textit{interventional} causal representation learning~\cite{scholkopf2021toward, ahuja2023interventional}, which considers scenarios in which actions (interventions) are applied into a system, changing its properties and facilitating better representations. In this sense, one could leverage such strategies to study the impact of changes in experimental conditions to both understand their influence into the system and better extract the underlying physical model~\cite{nautrup2022operationally, whitney2019dynamics}.

The results of our study also offer practical implications for model simplification and computational efficiency. In phenomenological models, characterized by multiple input parameters, the reduction of the dimensionality of the parameter space can significantly decrease the computational cost associated with the modeling and simulation of stochastic processes, thus enabling more efficient analysis and prediction of their behavior. Thus, the proposed approach offers a promising avenue for advancing the modeling and analysis of stochastic systems, enabling researchers to gain deeper insights into physical processes.


\begin{acknowledgments}
G.M.-G. acknowledges funding from the European Union.
C.M.  acknowledges support through Grant No. RYC-2015-17896 funded by MCIN/AEI/10.13039/501100011033 and ``ESF Investing in your future'', Grants No. BFU2017-85693-R and No. PID2021-125386NB-I00 funded by MCIN/AEI/10.13039/501100011033/ and ``ERDF A way of making Europe,'' and Grant AGAUR 2017SGR940 funded by the Generalitat de Catalunya.
G.F.-F., A.D., and M.L. acknowledge support from: 
European Research Council AdG NOQIA; 
MCIN/AEI (PGC2018-0910.13039/501100011033,
CEX2019-000910-S/10.13039/501100011033,
Plan National FIDEUA PID2019-106901GB-I00, 
Plan National STAMEENA PID2022-139099NB, I00,
project funded by
MCIN/AEI/10.13039/501100011033 and by the ``European
Union NextGenerationEU/PRT'' (PRTR-C17.I1), FPI);
QUANTERA MAQS PCI2019-111828-2); QUANTERA
DYNAMITE PCI2022-132919, QuantERA II Programme
co-funded by European Union’s Horizon 2020 program
under Grant Agreement No. 101017733); Ministry for
Digital Transformation and of Civil Service of the Spanish
Government through the QUANTUM ENIA project call
- Quantum Spain project, and by the European Union
through the Recovery, Transformation and Resilience Plan
- NextGenerationEU within the framework of the Digital
Spain 2026 Agenda; Fundació Cellex; Fundació Mir-Puig;
Generalitat de Catalunya (European Social Fund FEDER
and CERCA program, AGAUR Grant No. 2021 SGR
01452, QuantumCAT\textbackslash U16-011424, co-funded by ERDF
Operational Program of Catalonia 2014-2020); Barcelona
Supercomputing Center MareNostrum (FI-2023-3-0024);
(HORIZON-CL4-2022-QUANTUM-02-SGA PASQuanS2.1,
101113690, EU Horizon 2020 FET-OPEN OPTOlogic,
Grant No. 899794), EU Horizon Europe Program (This
project has received funding from the European Union’s
Horizon Europe research and innovation program under
Grant Agreement No. 101080086 NeQST Grant Agreement
101080086 NeQST); ICFO Internal ``QuantumGaudi''
project; European Union’s Horizon 2020 program under the
Marie Sk\l odowska-Curie Grant Agreement No. 847648; ``La
Caixa'' Junior Leaders fellowships, ``La Caixa'' Foundation
(ID 100010434): CF/BQ/PR23/11980043. Funded by the
European Union. Views and opinions expressed are, however,
those of the author(s) only and do not necessarily reflect those
of the European Union, European Commission, European
Climate, Infrastructure and Environment Executive Agency
(CINEA), the European Research Executive Agency nor any
other granting authority. Neither the European Union nor any
granting authority can be held responsible for them.
\end{acknowledgments}

\makeatletter
\def\bibsection{\section*{\refname}} 
\makeatother

\bibliography{ms.bib}

\appendix

\section{Machine learning pipeline}
In this section, we provide an overview of the machine learning architecture used in the main text, along with further explanations on the loss function, datasets, and training.
A Python implementation of the developed software, mainly based on the PyTorch~\cite{paszke2019pytorch} and fastai~\cite{howard2020fastai} libraries, is provided in~\cite{fernandez2023spivae}.

\subsection{Architecture}\label{sec:arch}
The machine learning model used throughout our work is schematically represented in~\cref{fig:arch}, and its layers' parameters are specified in \cref{tab:arch_details}.
The inputs to the model are the displacements of a $d$-dimensional trajectory of length $T$, represented as a tensor of size $d\times (T-1)$.
The model is able to produce the displacements of a new trajectory with arbitrary length $T'$, as we extend below. This work considers a generative model whose output is not directly the trajectory's displacements but rather their probability distribution.
To model these distributions, we use a Gaussian distribution $\mathcal{N}(\mu_t,\sigma_t)$ for each displacement $\Delta x_t$. 
Hence, the network here predicts $n=2$ parameters, $\mu_t$ and $\sigma_t$, and then $\Delta x_t$ is sampled from the associated distribution.
In the present work, we focus on one-dimensional trajectories ($d=1$) and consider a fixed trajectory length of $T=400$ for all trainings. 
Nonetheless, the model is length independent, as shown in \cref{fig:generation},
where we generate trajectories of 6000 time steps from a pre-trained model with  $T=400$.

\begin{table}[htbp]
\begin{tabular}{ll}
\hline\hline
Layer type                            & Output size                  \\ \hline
Input                                 & $B \times d \times (T-1)$    \\[2ex]
\multicolumn{2}{l}{Encoder}                                          \\[1ex]
\ \ $4\ \times$ 1D Conv. ($N_c=16$)        & $B \times 16 \times (T-9)$   \\
\ \ 1D adaptive (avg + max) pooling       & $B \times 16 \times (16+16)$ \\
\ \ Flatten                               & $B \times 512$               \\
\ \ MLP (200 and 100 neurons)             & $B \times 100$               \\[2ex]
Latent distribution                   & $B \times 12$                \\
Latent layer ($|\mathbf{z}|=6$)       & $B \times 6$                 \\[2ex]
\multicolumn{2}{l}{Decoder}                                          \\[1ex]
\ \ MLP (100, 200, and 512 neurons)           & $B \times 512$               \\
\ \ Reshape                               & $B \times 16 \times 32$      \\
\ \ Interpolation                         & $B \times 16 \times (T-9)$   \\
\ \ $3\ \times$ 1D transposed Conv. ($N_c=16$)              & $B \times 16 \times (T-1)$     \\
\ \ 1D transposed Conv.  ($N_c=|\mathbf{z}|$)         & $B \times \ \, 6 \times (T-1)$      \\
\ \ WaveNet ($d_c =[1, 2, 4, 8], \mathrm{RF} = 32$) & $B \times nd \times (T\!-\!1\!-\!\mathrm{RF})$ \\[2ex]
Sampled output                                & $B \times d \times T'$ \\
\hline\hline
\end{tabular}
\caption{\textbf{Architecture layers' details.} We use ReLU as non-linear activation function in all layers, kernel of size three and stride one on the convolutional (Conv.) layers, and no padding. We abbreviate the terms: batch size $B$, dilation $d_c$, number of convolutional channels $N_c$, and receptive field RF.}
\label{tab:arch_details}
\end{table}

The core of the model is inspired by~\cite{chorowski2019unsupervised} and consists of a convolutional variational autoencoder~(VAE) with an autoregressive~(AR) decoder.
The architecture, as any VAE-like structure, has three main components: 
i) a convolutional encoder that compresses the input into the latent neurons;
ii) a set of \textit{probabilistic} latent neurons;
iii) a decoder that up-samples the latent representation to control the generation of new outputs.

\paragraph*{i) Encoder --} As presented in \cref{tab:arch_details}, the encoder consists of a stack of four convolutional layers, followed by an adaptive (average and maximum) pooling layer, and a two layered multilayer perceptron (MLP) that transforms the data into the appropriate latent dimension.

\paragraph*{ii) Latent space --} Following the typical VAE construction~\cite{kingma2013auto}, the latent space consists of a set of probabilistic neurons of size $|\mathbf{z}|$.
Throughout this work, we consider $|\mathbf{z}|=6$.
To facilitate training, we consider the widely known \textit{reparameterization trick}~\cite{kingma2013auto}:
instead of considering a probabilistic neuron,
we sample it from two, each representing the mean and variance of a Gaussian,
while externalizing the noise.
This way, one can properly backpropagate the error through this layer.

\paragraph*{iii) Decoder --} The decoder consists of two distinct parts. First, a convolutional module upsamples the latent vector to a higher dimensional space. As presented in \cref{tab:arch_details}, this is done by reversing the encoder modules.
That is, stacking various layers MLP, interpolation, and transposed convolution layers.
Second, an autoregressive module, based on the WaveNet architecture~\cite{oord2016wavenet}, generates the model's output.
This kind of networks uses a finite number of previous data points, defined as the receptive field (RF), to predict the forthcoming one. In the current work, $\mathrm{RF}=32$. 
During training, as shown in \cref{fig:arch}, the AR module receives as input the trajectory's displacements.
The model uses an amount of initial displacements corresponding to the RF to make the first prediction.
Thus, the prediction length is fixed to $T-1-\mathrm{RF}$.
Our experiments show that padding the input with zeros inevitably creates artifacts, as the padding lacks the trajectory features.
When performing inference (i.e., generating new trajectories), WaveNet generates new outputs (i.e., the displacement $\Delta x_t$) by recursively feeding its own previous outputs as input (i.e., the displacements $\Delta x_{t-1}, \Delta x_{t-2}, ...$).
By means of this recursive sampling, the generated trajectories can have an arbitrary length $T'$. Importantly, in all cases, the upsampled latent vector generated by the convolutional module is fed as a conditioning to each layer of the WaveNet.

\subsection{Loss function}\label{sec:loss}
Below, we discuss some key aspects of the loss function of the model which are relevant to understanding the loss distribution of the datasets used in our study.
As a $\beta$-VAE model, the proposed model has a loss consisting of two components, which can be expressed as follows:
\begin{equation}\label{eq:loss}
    \mathcal{L}=
    - \log p_{\theta}(\mathbf{\Delta x}|\mathbf{z})
    \ + \ 
    \beta\ D_{KL}(p_\phi(\mathbf{z}) \parallel g(\mathbf{z})).
\end{equation}
The first term is a reconstruction loss that evaluates the similarity between the inputs and the model's outputs.
Here, we employ for that the negative log-likelihood (NLL), which is asymptotically equivalent to the Kullback-Leibler divergence ($D_{KL}$) of the predicted distribution from the data's true probability distribution. 
The second term measures the similarity between the distribution of the latent neurons $p_\phi(\mathbf{z})$ and their prior $g(\mathbf{z})$.
In the following,
we focus on the reconstruction term in the AR framework and
refer the reader to~\cite{higgins2016beta}
for a study on the role of $\beta$ weighting the second term.

The AR formalism considers that the input data distribution can be described as the product of conditional probabilities 
\(p(\mathbf{x})=\prod_t p(x_t|\mathbf{x}_{<t})\),
where each conditional probability is a function of the previous values in the data with some ordering $<t$, i.e., $\mathbf{x}_{<t}=x_{1}, ..., x_{t-1}$. 
In practice, one introduces a receptive field (RF), such that $p(x_t|\mathbf{x}_{<t})\rightarrow p(x_t|x_{t-\mathrm{RF}}, ..., x_{t-1})$. This term is of high importance, as it defines the amount of previous information considered when recursively predicting next steps.
In AR models, each of these conditional probabilities is modeled as a parameterized probability distribution
\(p(x_t|\mathbf{x}_{<t})\approx p_\theta(x_t|\mathbf{x}_{<t})\) whose parameters $\theta$ can be found by maximizing a likelihood function w.r.t. the input data.
As commented above, we consider here that $p_\theta(\Delta x_t|\mathbf{\Delta x}_{<t}) = \mathcal{N}(\mu_t, \sigma^2_t)$, where $\mu_t$ and $\sigma_t$ are calculated by the AR model considering all $\mathbf{\Delta x}_{<t}$ as defined by the recursive scheme mentioned above. We emphasize that $\mu_t$ and $\sigma_t$ are computed by the decoder and hence are a function of $z$. In the equations, this dependency is intentionally omitted for simplicity and ease of understanding. Then, the parameters $\theta$ of the network are optimized by minimizing the NLL, namely
\begin{equation}
- \log p_{\theta}(\mathbf{\Delta x}|\mathbf{z})
=-\sum\limits_{i=1}^{N}\sum\limits^{T-1}_{t=1} \log\mathcal{N}\left(\mu_t^{(i)}, \sigma^{2\,(i)}_t\right),
\end{equation}
where $N$ is the total number of samples in the training dataset and $T-1$ is the total number of considered displacements.
The NLL for a single displacement and time step reads 
\begin{equation}
-\log\mathcal{N}(\mu_t, \sigma_t^2)=  \log\left(\sigma_t\sqrt{2\pi}\right)     +\frac{(\Delta x_t-\mu_t)^2}{2\sigma_t^2},
\end{equation}
where the minimum loss is achieved when the input displacement $\Delta x_t$ coincides with the predicted $\mu_t$.
Once this is achieved, decreasing $\sigma_t^2$ effectively lowers the NLL.

\begin{figure}[htbp]
    \centering
    \includegraphics[width=\linewidth]{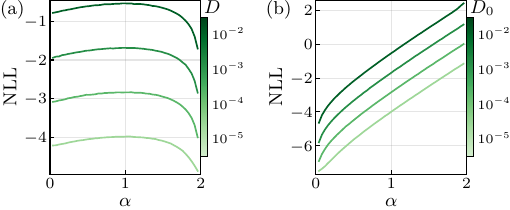}
    \caption{\textbf{Relation between the loss and the diffusion parameters.} Negative log-likelihood (NLL) of two models, trained with: (a) FBM and (b) SBM trajectories as a function of the trajectories' anomalous exponent $\alpha$. Colors encode the diffusion coefficient $D$ and $D_0$ for FBM and SBM respectively.}
    \label{fig:NLL}
\end{figure}

In this work, we consider data generated via Gaussian processes. Taking Brownian motion (BM) as an example, it can be seen that, in a properly trained model, the variance $\sigma_t^2$ defined above must be related to the variance of the input's displacements, which are directly connected to the diffusion coefficient via $\sigma^2=2D\Delta t$, where $\Delta t = 1$ for all cases considered in this work.
This means that, in the perfect training scenario, for BM, $\mu_t = 0$ and $\sigma_t^2 = 2D$ $\forall t$.
As a lower variance $\sigma_t^2$ implies a lower NLL, trajectories with lower diffusion coefficient will have lower NLLs.
This not only affects BM trajectories, but also FBM and SBM ones. In those cases, we expect that $\mu_t \neq 0$ for most $t$, but still lower diffusion coefficients will imply lower variances $\sigma_t^2$.
We observe this behavior in both FBM and SBM datasets, as shown in \cref{fig:NLL}.
Moreover, the aging effect, present in SBM, implies that the diffusion coefficient scales as $D_\alpha(t)=\alpha D_{0} t^{\alpha-1}$. Hence, a larger $\alpha$ implies a larger $D(t > 1)$ for a given $D_0$. Such effect is also seen in \cref{fig:NLL}, where the NLL increases for larger $\alpha$. Oppositely, for FBM, the NLL remains almost constant in the whole $\alpha$ range. 

\subsection{Datasets}\label{sec:datasets}
BM, FBM, SBM, and BM with random confinement trajectories are generated via the \texttt{andi\_datasets} package~\cite{munoz2023andichallenge}.
Each dataset consists of about \(100\, 000\) trajectories of
total time length \(T=400\), with 10 diffusion coefficients spaced logarithmically 
$D \in [10^{-5}, 10^{-2}]$. The same range is considered for $D_0$
in the SBM trajectories. 
Moreover, both FBM and SBM datasets consider 21 anomalous exponents equally spaced
in the range $\alpha \in [0.2, 1.8]$, for which we then consider 476 trajectories for each of the parameters’ combination.
In the case of BM with random confinement, we generate a dataset with $\alpha=1$, 21 radii of confinement $r\in [0.005, 0.05]$, the same 10 diffusion coefficients of FBM, and boundary compartment transmittance of $1/20$.

We split the training and validation  datasets following the standard 80\% and 20\% proportion, respectively.
In most of our analysis, the range of the test data was within the training range.
However, for \cref{fig:latent}(b) and \cref{fig:generation}(c), we used an extended test dataset with 49 equally spaced values of $\alpha\in [0.04, 1.96]$.
Similarly, for \cref{fig:latent}(a,c), we used an extended test dataset that spanned 49 $D\in [3.16\cdot 10^{-6}, 3.16\cdot 10^{-2}]$.

\paragraph*{Anomalous exponent estimation ---} In order to estimate the anomalous exponent of trajectories generated by the model, we use different averages to the mean squared displacement (MSD). For the trajectories generated with the model trained on FBM trajectories, we estimate $\alpha_g$ by fitting the time average mean squared displacement (TA-MSD). Considering that the trajectory is sampled at $T$ discrete times $t_i=i\Delta t$,
\begin{equation}
    \mathrm{TA\!-\!MSD} (\Delta t) = \frac{1}{T-\Delta t}\sum_{i=1}^{T-\Delta t} [x(t_i)-x(t_i+\Delta t)]^2,
\end{equation}
where $\Delta t$ is the time lag.
In \cref{fig:generation}(b),
we fit the TA-MSD with a sliding window of three time lags, ranging from the smallest time lag up to $\Delta t = 420$.
For the insets, we perform a linear fit of the TA-MSD on the highlighted range of time lags, respectively from $\Delta t = 1$ to 20, and from $\Delta t = 380$ to 400.
The same method would not work for SBM due to its weakly non-ergodic nature.
Hence, $\alpha_g$ shown in \cref{fig:generation}(c) is estimated by fitting the time and ensemble-average mean squared displacement (TEA-MSD).
The TEA-MSD for a fixed time lag can be defined in terms of the TA-MSD for the $i$-th trajectory as:
\begin{equation}
    \mathrm{TEA\!-\!MSD} (\Delta t) = \frac{1}{N}\sum_{i=1}^{N} \mathrm{TA\!-\!MSD}(\Delta t)_i,
\end{equation}
where $N$ is the total number of generated trajectories.
We take the last 55\% of the trajectory length and $\Delta t=2$ to assure statistical significance.

\subsection{Training}\label{sec:training}
In this section,
we specify the setup used for training our model, including the initialization of the model's parameters and the set of hyperparameters used during the training.

To minimize the loss, we updated the model's parameters using the Adam optimizer~\cite{kingma2014adam} with a maximum learning rate selected using the learning rate finder tool from the fastai library~\cite{howard2020fastai}, usually found around $10^{-4}$.
Additionally, we scheduled both the learning rate and the optimizer's momentum using the one cycle policy from~\cite{smith2018disciplined}.
The batch size was set to 256 trajectories.

We found that a proper initialization was crucial to obtain good results.
We used Kaiming He's initialization~\cite{he2015delving} with fan out mode, except for the latent neurons representing the logarithm of the variance, which were initially set to zero to prevent overflow in the initial stages.

To ensure proper learning when using an autoregressive model, it is important to use an annealing schedule of $\beta$~\cite{bowman2015generating}. In this study, we first train until convergence with $\beta=0$, and then, we follow a monotonically increasing annealing schedule for $\beta$ to minimize the number of informative neurons while having a good reconstruction loss.
We found a good compromise of $\beta$ on the order of $5\cdot 10^{-3}$.

\section{Latent neurons encode stochastic parameters}\label{sec:3d_latent}
\begin{figure}[htbp]
    \centering
    \includegraphics[width=\linewidth]{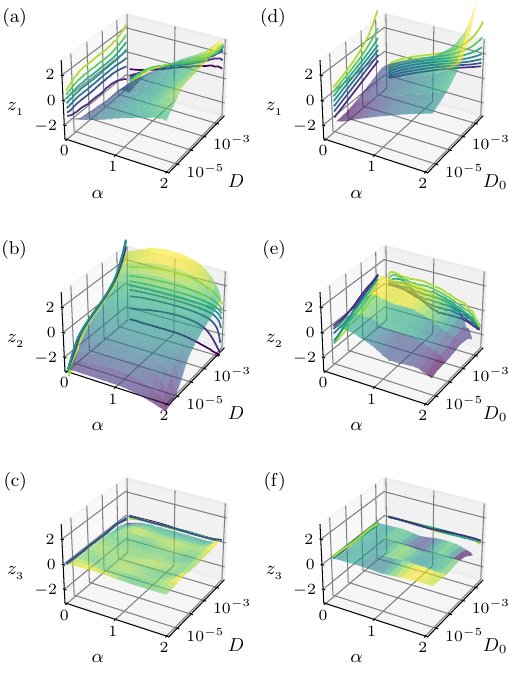}%
    \caption{\label{fig:VAEWN_latent_aD_3d}
    \textbf{3D representation of the latent space.} 
    Distribution of latent neurons activations \(z_i\) w.r.t. the anomalous exponent $\alpha$ and the diffusion coefficient,  $D$ for FBM (left column) and $D_0$ for SBM (right column). $z_1$ and $z_2$ show a clear relation w.r.t. to the diffusive parameters, while $z_3$ has been noised out and is completely uninformative.}
\end{figure}

\Cref{fig:VAEWN_latent_aD_3d} shows a more detailed representation of the learned latent space. The 2D projections of these figures are shown in the main text (\cref{fig:latent}). As shown, the model learns a combination of both relevant parameters in the latent neurons.
For FBM, \cref{fig:VAEWN_latent_aD_3d}(a,b), 
the latent neurons \(z_1\) and \(z_2\) encode both
the anomalous exponent \(\alpha\) and the diffusion coefficient \(D\). 
This encoding is smooth, continuous, and generally independent except for big $\alpha$ that shows how the same neuron is encoding a non-linear relationship of the parameters.
In contrast, as illustrated in the main text, for SBM the model learns two distinct combinations of the parameters,  
\(\log \alpha D_0 T^{\alpha-1}\) and $\log \alpha D_0 T^{-(\alpha-1)}$.
These combinations can be considered as complementary non-linear transformations of $\alpha$ and $D_0$. These relationships are depicted in \cref{fig:VAEWN_latent_aD_3d}(d,e), 
where the latent neurons directly represent the combination of both $\alpha$ and $D_0$. In both datasets, all but two latent neurons are noised out and show no relationship with the parameters, as depicted in
\cref{fig:VAEWN_latent_aD_3d}(c,f).

Changes

2026-01-16 Corrected sign of second term in \cref{eq:loss}.
\vfill


\end{document}